\documentclass{article}
\usepackage{ijcai25}

\usepackage{times}
\usepackage{soul}
\usepackage{url}
\usepackage[hidelinks]{hyperref}
\usepackage[utf8]{inputenc}
\usepackage[small]{caption}
\usepackage{graphicx}
\usepackage{amsmath}
\usepackage{amsthm}
\usepackage{booktabs}
\usepackage{algorithm}
\usepackage{algorithmic}
\usepackage[switch]{lineno}

\usepackage{multirow}
\usepackage[table]{xcolor} 
\usepackage{array}
\usepackage[caption=false]{subfig}
\usepackage{textcomp}
\usepackage{stfloats}
\usepackage{verbatim}
\usepackage{newclude}
\usepackage{pifont}
\usepackage{xcolor}
\usepackage{listings}
\usepackage[capitalise]{cleveref}

\newcommand{\tool}{\textit{Hoyen}}
\newcommand{\toollink}{\textcolor{blue}{https://hoyen.tjunsl.com/}}
\def\eg{\emph{e.g.}} 
\def\ie{\emph{i.e.}}

\def\etal{\emph{et al.}}

\title{Unlocking User-oriented Pages: Intention-driven Black-box Scanner for Real-world Web Applications}

\author{
Weizhe Wang$^1$\and
Yao Zhang$^1$\footnote{Corresponding author}\and
Kaitai Liang$^2$\and
Guangquan Xu$^{1*}$\and
Hongpeng Bai$^1$\and
Qingyang Yan$^1$\and
Xi Zheng$^4$\And
Bin Wu$^1$\\
\affiliations
$^1$Tianjin University\\
$^2$Delft University of Technology\\
$^3$Macquarie University\\
}

\begin{document}

\maketitle

\begin{abstract}
Black-box scanners have played a significant role in detecting vulnerabilities for web applications. A key focus in current black-box scanning is increasing test coverage (\ie, accessing more web pages). However, since many web applications are user-oriented, some deep pages can only be accessed through complex user interactions, which are difficult to reach by existing black-box scanners.
To fill this gap, a key insight is that web pages contain a wealth of semantic information that can aid in understanding potential user intention. Based on this insight, we propose \tool, a black-box scanner that uses the Large Language Model to predict user intention and provide guidance for expanding the scanning scope.
\tool\, has been rigorously evaluated on 12 popular open-source web applications and compared with 6 representative tools. The results demonstrate that \tool\, performs a comprehensive exploration of web applications, expanding the attack surface while achieving about $2\times$ than the coverage of other scanners on average, with high request accuracy. Furthermore, \tool\, detected over 90\% of its requests towards the core functionality of the application, detecting more vulnerabilities than other scanners, including unique vulnerabilities in well-known web applications. Our data/code is available at \toollink
\end{abstract}

\section{Introduction}

Web applications have become a crucial component of the modern Internet. Detecting vulnerability in web applications is significant for maintaining cybersecurity.
Due to the dynamic and complex nature of web applications, their source code is often opaque to users, making it difficult to access directly. Therefore,
Black-box detection for Web applications have gained significant attention in recent years due to their ability to test without relying on specific systems or code types. 

Black-box testing for web applications has been studied for many years, with a major focus on how to enhance coverage. In early research, most of the approaches rely on breadth-first search (BFS)
and random navigation strategies 
to explore target web applications \cite{doupe2012enemy,eriksson2021black,pellegrino2015jak}.
Among them, \cite{doupe2012enemy} identified vulnerabilities by generating user input vectors based on inferred state machines. However, their heuristic-based approach faces challenges in managing complex state transitions. In light of this, \cite{eriksson2021black} built on \cite{pellegrino2015jak} designed a data-driven crawler to model navigation and track inter-state dependencies.
However, these methods are suboptimal for thoroughly testing web applications, as the absence of specific knowledge about the target. Consequently, they may inadvertently trigger unintended functions or actions that should remain inactive (such as logging out), leading to premature or unintended test terminations and making it difficult to detect deeper and more complex vulnerabilities.

In recent years, machine learning is extensively utilized in web application testing to optimize the efficacy of web application scanners \cite{hannousse2024twenty}. 
Traditional approaches based on machine learning \cite{melicher2021towards} to detect vulnerabilities suffer from suboptimal cost-effectiveness. These approaches rely heavily on customized pre-trained models, making it difficult to dynamically adapt to the diverse and evolving nature of web applications.
The introduction of the Transformer framework \cite{vaswani2017attention} has led to the rise of numerous Large Language Models (LLMs).
LLMs are distinguished by their powerful natural language processing and generalized reasoning abilities, consistently demonstrating superior performance in recognition and inference tasks in textual and multimodal domains \cite{you2024ferret}. A substantial body of research has exploited the advanced capabilities of LLMs to identify software vulnerabilities and system threats \cite{zheng2024testing,jiang2024fuzzing,xia2024fuzz4all,oliinyk2024fuzzing,wang2024llmif}. However, the ability of LLMs to detect vulnerabilities in web applications remains constrained due to the immense scale of web applications and the complex interrelations between their various states. 

\begin{figure*}[!htbp]
    \centering
    \includegraphics[width=0.95\linewidth]{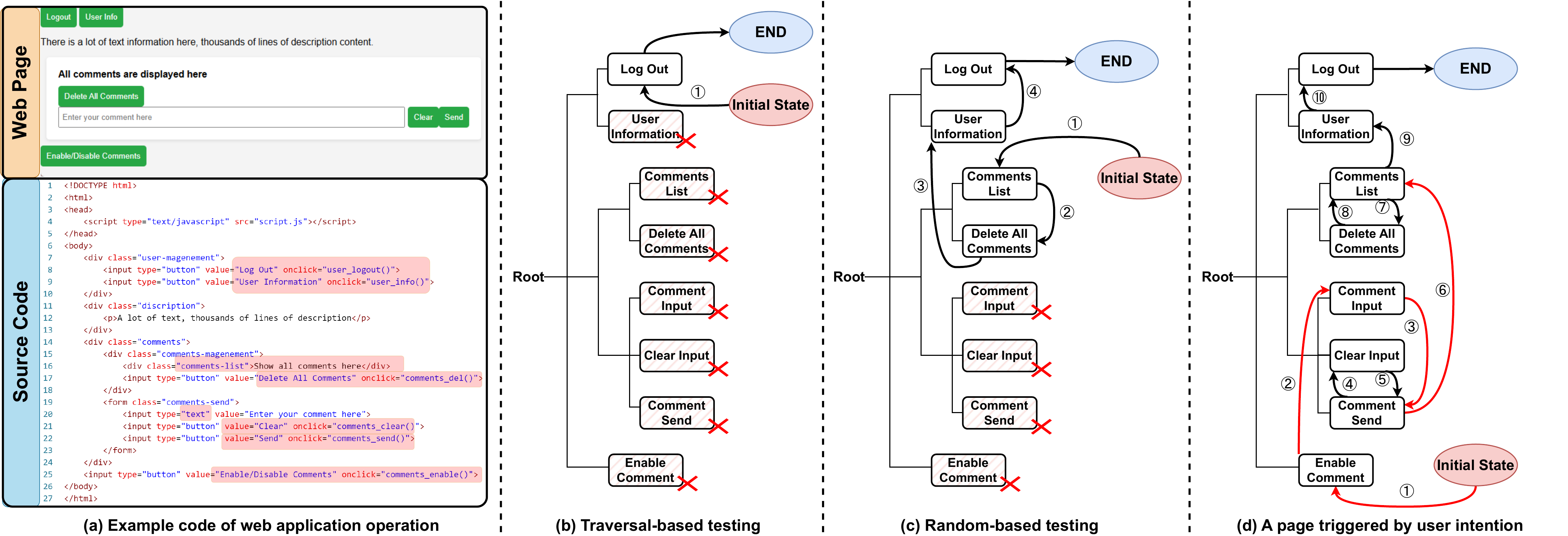}
    \caption{Motivation example. The four sub-diagrams represent the web application code with interface content, and request flow for the traversal, random, and proposed strategies. The red line indicates the flow of the actual test that triggers the vulnerability.}
    \label{fig:motivation_compare}
\end{figure*}

Although existing researchers have made efforts to improve the coverage of black-box web testing, they have yet to consider it from the perspective of user interaction. In fact, most contemporary web applications are typically designed around user interactions, which makes many deep-level web pages can only be reached through specific user operations. These pages pose a significant hurdle for existing black-box testing methods. We take Figure~\ref{fig:motivation_compare} as an example, where Figure~\ref{fig:motivation_compare}-a shows a web page along with its source code, illustrating the commenting functionality of the web page with several functions, \eg, ``Comment Input", ``Comment Send", ``Clear Input", \etal Testing the commenting functionality requires enabling the comment feature, inputting a comment, and submitting it, and the comment should then appear in the list. It is crucial to avoid actions like ``log out" or ``delete all comments", which can disrupt the testing process. We take two state-of-the-art tools, \ie, \cite{xray} and \cite{eriksson2021black}, as examples to show why existing works struggle in this case. As a traversal-based tool, \cite{xray} will always trigger ``log out" as soon as it begins to explore the web app, failing to reach other functions (step \ding{172} in Figure~\ref{fig:motivation_compare}-b). \cite{eriksson2021black} employs a randomized strategy, allowing it to test more functions. However, due to the lack of contextual understanding, it fails to prioritize enabling the commenting functionality, preventing comment submission for testing. It also prematurely triggers actions like comment deletion and logout  (\ding{172} to \ding{175} in Figure~\ref{fig:motivation_compare}-c), causing early test termination. Fortunately, as illustrated in the code of Figure~\ref{fig:motivation_compare}-a, web pages contain rich semantic information (highlighted in red) that can assist in predicting user intent. Figure~\ref{fig:motivation_compare} shows how a user interacts with the comment function. Specifically, the user first inputs a comment (\ding{172} to \ding{173}), then either clears the input (\ding{174}) or sends it to trigger a deeper page (\ding{175} to \ding{181}). Alternatively, the user may intend to delete the comment after it appears in the comment list (\ding{178} to \ding{179}). Clearly, understanding user intent can help identify the path to the deeper page.

From the above example, it can be seen that semantic information in web pages can infer potential user intentions, thereby helping to identify in-depth pages. In recent years, advancements in LLMs have opened new opportunities for extracting semantic data from code. Consequently, a natural idea is to leverage LLMs to analyze semantic information from web pages. Nevertheless, there are still some challenges that need to be addressed.
\textbf{C1: The analysis for large-scale web page.} Real-world web applications always contain a vast amount fo content. Dealing with the web page with thousands of lines code presents a significant challenge for LLMs, which struggle with token limitation and performance degradation over extended content. Besides, the vast amount of content can also make it difficult for large models to focus on key information, leading to misjudgments by the models.
\textbf{C2: Intricate dependencies in real-world scenarios.} In real-world web applications, interdependencies can significantly impact the application's state~\cite{doupe2012enemy,eriksson2021black}. For instance, in Figure~\ref{fig:motivation_compare}-c, the failure to recognize the dependency between enabling and the ``Comment Send" action led to triggering the wrong action prematurely. As web applications continue to grow in size, the dependencies become increasingly complex, making their analysis more challenging.

To mitigate above challenges, we introduce an intention-driven black-box scanner for web applications, named \tool. Specifically, 
to address \textbf{C1}, we extract semantically relevant information from web pages and analyze the structure information between the page elements, ensuring efficient page analysis by LLM.
To address \textbf{C2}, we leverage LLM, understanding the intention path and element context based on the real user intention, to infer operations that follow the user's intent and handle dependencies in the application.

In summary, this paper makes the following contributions.

\begin{itemize}
    \item We introduced \tool, a novel black-box scanner for web applications that employs contextual semantics and content awareness to test web applications.
    \item We proposed a semantic and content-awareness-based user intention prediction approach to address the complexity of web application dependencies and the limitations of conventional access strategies in exploring deeper application pages.
    Additionally, we proposed a function-based page content refinement approach to extract key semantics and understand the extensive information of large-scale web application pages. 
    \item We conducted a comprehensive comparative analysis of \tool\, versus 6 other academic and industrial scanners in 12 different web applications. Our experimental results demonstrate that \tool\, exhibits superior page coverage and request accuracy compared to other methods while uncovering more vulnerabilities.
\end{itemize}

\section{Related Work}

\begin{figure*}[!htbp]
    \centering
    \includegraphics[width=0.95\linewidth]{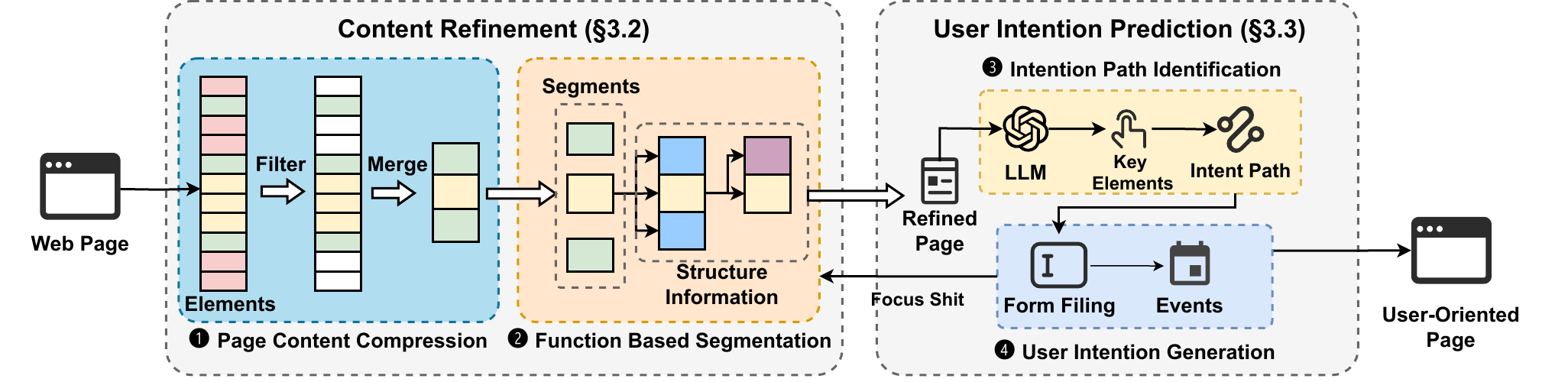}
    \caption{The framework of \tool}
    \label{fig:framework}
\end{figure*}

\cite{doupe2012enemy} proposed Enemy of the State, a black-box scanner that identifies security vulnerabilities by inferring state machines but struggles to handle complex states. To address challenges posed by JavaScript-based applications, \cite{pellegrino2015jak} introduced j{\"a}k, a dynamic crawler leveraging client-side JavaScript analysis. Building on this, \cite{eriksson2021black} developed BlackWidow, a data-driven scanner that enhances code coverage but remains susceptible to state loss due to incorrect edge traversal. \cite{zheng2021automatic} and \cite{zhang2022unirltest} applied curiosity-driven reinforcement learning to enable efficient and adaptive exploration for automatic web testing. Still, their approaches lacked semantic understanding, limiting their ability to capture interdependency correlations within web applications. Meanwhile, \cite{kirchner2024dancer} focused on blind XSS detection using a polyglot methodology but failed to account for broader semantic relationships.

In contrast with black-box testing, white-box testing with direct access to source code, allows deeper analysis and vulnerability detection, such as through machine learning for XSS detection \cite{kaur2023detection} and taint analysis \cite{luo2022tchecker,guler2024atropos,fioraldi2020afl++}, which has proven effective in identifying vulnerabilities. However white-box testing is often impractical for many systems due to language dependencies and strict requirements. Gray-box testing offers partial system insight and balances these approaches \cite{olsson2024spider}. 
Although white-box and gray-box testing offer some advantages in the detection of vulnerabilities in web applications, black-box testing remains the preferred method due to its wide applicability \cite{trickel2023toss,al2023sqirl,zhao2023remote}. However, the complexity of current black-box approaches limits the widespread of AI techniques in black-box vulnerability detection for web applications.

\section{Approach}\label{section:Approach}
\subsection{Overview}

Black-box testing is a widely used method for identifying vulnerabilities in web applications. Its main advantage is that it does not necessitate a comprehensive understanding of the target system, eliminating the need for system-specific adaptations. Thus, it has garnered considerable attention and adoption in academic and industrial contexts.

Figure~\ref{fig:framework} shows the overview of \tool, which consists of two component: Content Refinement (Section~\ref{section:pageCompExtr}) and User intention Prediction (Section~\ref{sec:InterntionPrediction}). 

 \textbf{Content Refinement}. Given a real-world web page, \tool~first compress the page to eliminate the elements which irrelevant to semantics and merge the similar elements to reduce the content scale (See \ding{182}). Then \tool~segments the web page into specific functional block to provide structure information of pages for LLM (See \ding{183}). 
 
 \textbf{User Intention Prediction}. After refining the page content, \tool~identify the key elements associate with page action to infer the possible user intent (see \ding{184}). Then, \tool~generates user intentions by understanding the dependence between different elements (see \ding{185}).
 \subsection{Content Refinement}\label{section:pageCompExtr}
 
For large real-world web applications, we first reduce the content of the page. This section consists of two parts, \ie, page content compression and function based segmentation.

\subsubsection{Page Content Compression}

To compress the page content, our key idea is to eliminate semantically irrelevant information and merge similar pages, thereby reducing the overall page size.

We begin by organizing all major HTML tags and functions, and classify elements into three categories based on their attributes: 1) \textit{Style elements}, used for front-end customization like CSS styles; 2) \textit{association elements}, which are semantically related consecutive elements, such as \verb|<p>| and \verb|<br>|; and 3) \textit{core functional elements}, supporting core application functions, such as forms. Since style elements are irrelevant to semantic understanding, we use a matching-based method to traverse and remove them. Association elements, such as continuous long paragraphs of text content, which often appear in sequence with similar semantics, are processed by focusing solely on their value and content, then merging and abbreviating their content with semantics. For core functional elements, we recursively examine their content and child elements, eliminate irrelevant content, retaining only relevant functional and semantic information.

After eliminating irrelevant information from the pages, we merge similar pages that may widely exist in the application. To this end, we use a multidimensional similarity measure to evaluate their similarity.
We first check whether the URLs of two pages adhere to the same-origin policy. Based on this, we use the LCS algorithm to comprehensively measure the similarity of the URLs, request paths, and query parameters, ensuring they meet predefined similarity thresholds. Given the dynamic nature of page rendering, URL similarity alone cannot fully capture the similarity relationship between pages. Therefore, we also analyze page-specific HTML information. Since the DOM structure reflects functional features and the style indicates visual features, we evaluate their similarity comprehensively. Specifically, we use Tree Edit Distance to compare DOM structure similarity, and Jaccard Similarity for style comparison, and combine these measures with predefined weights for overall similarity. 

\subsubsection{Function Based Segmentation}\label{section:segmentation}
To help LLM understand the content, we segment the HTML pages based on functionality. The key insight comes from the hierarchical structure of the HTML DOM tree. Specifically, the related functions are generally grouped within the same branch, while adjacent functions and elements tend to be located at the same level of the tree. Parent-child relationships and branching patterns of the tree reflect the dependencies between elements.  Leveraging these characteristics, we are able to extract functionally and semantically relevant smallest blocks.  

\begin{figure}
    \centering
    \includegraphics[width=0.95\linewidth]{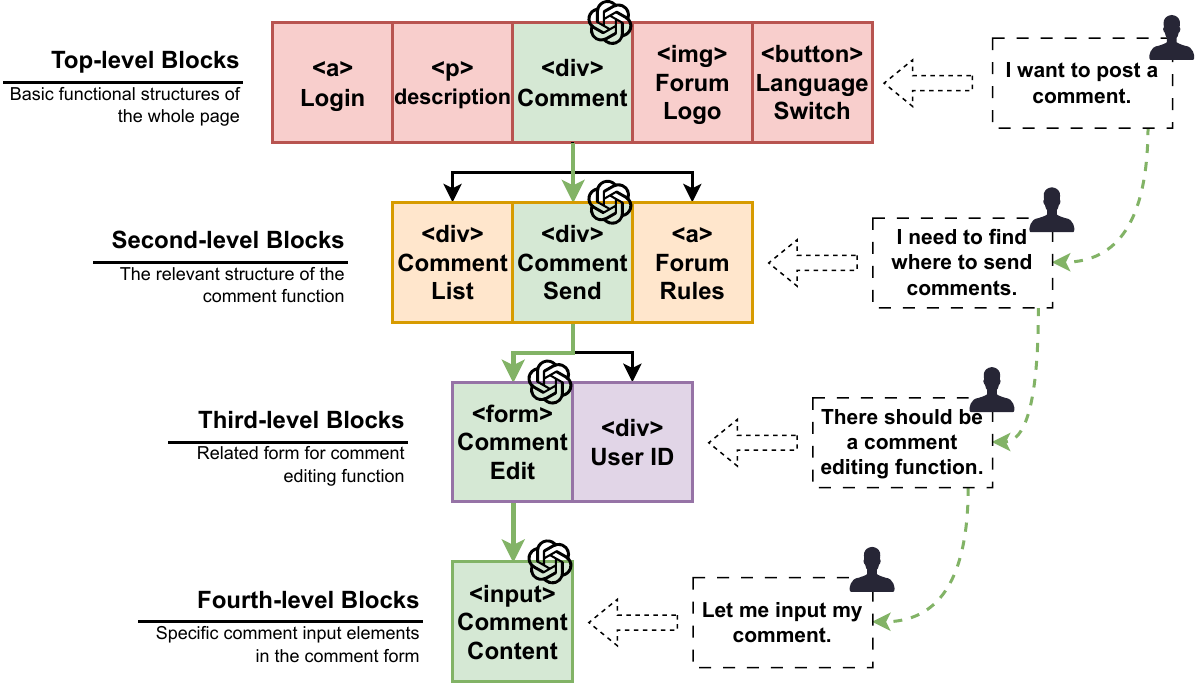}
    \caption{Example of function segmentation. Each lattice comprises structure tags and semantic information. This page is recursively segmented into multiple layers of structure, starting with the comment function, and culminating in the comment input element.}
    \label{fig:segmentation}
\end{figure}

We use the case shown in Figure~\ref{fig:segmentation} to illustrate the segmentation process. Starting from the top level of the HTML DOM tree, we divide the page into five functional blocks. For example, the block ``$<$div$>$ Comment" represents the comment function. If the scanner focuses on this block, it further segments it into smaller blocks based on their functions. In this case, the ``$<$div$>$ Comment" block is recursively divided into three sub-blocks, \ie, ``$<$div$>$ Comment list", ``$<$div$>$ Comment sent", and ``$<$div$>$ Rules", each of which serves a specific function. By recursively segmenting each level, we eventually obtain the smallest block containing semantic information, namely, ``$<$div$>$ Comment content". Thus we can get contextual information from top-level to low-level to help LLM better understand user intention.

\subsection{User intention Prediction}\label{sec:InterntionPrediction}

To complex dependencies in web applications, we propose user intention prediction to guide web application exploration. This section is divided into two parts, \ie, intention path identification and user intention generation.

\subsubsection{Intention Path Identification}

When a user selects a specific function in an application, their decision follows a certain logical thought process. On the application interface, the dependencies between elements create a sequence that reflects the user's intent. This sequence could form an intention path, guiding users toward accessing specific functions or user-oriented pages.

\begin{algorithm}[!ht]
\caption{Key Elements Identification}
\label{algo:elem_select}
\textbf{Input}: Web application page information, $Data\_page$ \\
\textbf{Output}: Key elements of the current page \\
\begin{algorithmic}[1]
\STATE $Page\_blocks \gets \texttt{page\_divided}(Data\_page)$

\REPEAT
    \STATE $Data\_blocks \gets \texttt{data\_extract}(Page\_blocks)$ \label{interalgo:es:data_extract}
    \STATE $Block\_select \gets \texttt{select\_block}(Data\_blocks)$ \label{interalgo:es:block_select}
    \IF {$\texttt{scale}(Page\_blocks) \leq \texttt{threshold}$}
        \STATE $Elems \gets \texttt{getValidElems}(Block\_select)$
    \ELSE
        \STATE $Page\_blocks \gets Block\_select$
    \ENDIF
\UNTIL $Elems \neq NULL$
\STATE $Elems\_Info \gets \texttt{get\_elems\_info}(Elems)$
\STATE $Elem\_key \gets \texttt{select\_elem}(Elems, Elems\_Info)$ \label{interalgo:es:elem_select}

\RETURN $Elem\_key$
\end{algorithmic}
\end{algorithm}
Our user intention path identification approach models the interaction and reasoning process of real users. Initially, we analyze the application structure and semantics from the function based segmentation approach (see Section~\ref{section:segmentation}). Based on this, we construct the prompt guided by the user intention and operation logic to drive LLM. By analyzing the overall context and the functional semantics of each block, we identify the block that best aligns with the user's intent. The user intention path is then constructed by combining the application's functional block, and progressively refining the intent from top to bottom. Through recursive traversal of the intention path and DOM structure, the scope is gradually narrowed, ultimately identifying the key elements that align with the user's intent for deep interaction and exploration within the web application. Taking Figure~\ref{fig:segmentation} as an example, by analyzing the semantics of the web application’s top-level blocks, the user may be potentially interested in the comment function and would like to post a comment. We then recursively examine the sub-structures, ultimately locating the specific comment input element, \ie the key element. Throughout this process, the LLM drives the user intention, constructing and identifying the intention path layer by layer along the DOM tree, enabling the precise execution of the comment posting function.

Our approach to identifying key elements is shown in the Algorithm~\ref{algo:elem_select}. First, we segment the page information based on function segmentation (Section~\ref{section:segmentation}) and extract data from specific segments $Page\_blocks$. Then, the LLM makes selection decisions for each block based on user intention (Line~\ref{interalgo:es:block_select}). The selected functional blocks of the DOM tree are used as inputs for subsequent extractions, and the user intention path is constructed through recursive looping until the selected block meets a predefined threshold (Less than 1/20 of the max LLM context token window size). At this point, relevant elements are extracted. Finally, by analyzing the semantics and context information of these elements, the LLM selects the key elements that align with the user's intention (Line~\ref{interalgo:es:elem_select}).

\subsubsection{User Intention Generation}

\begin{figure*}[!htbp]
    \centering
    \includegraphics[width=0.9\linewidth]{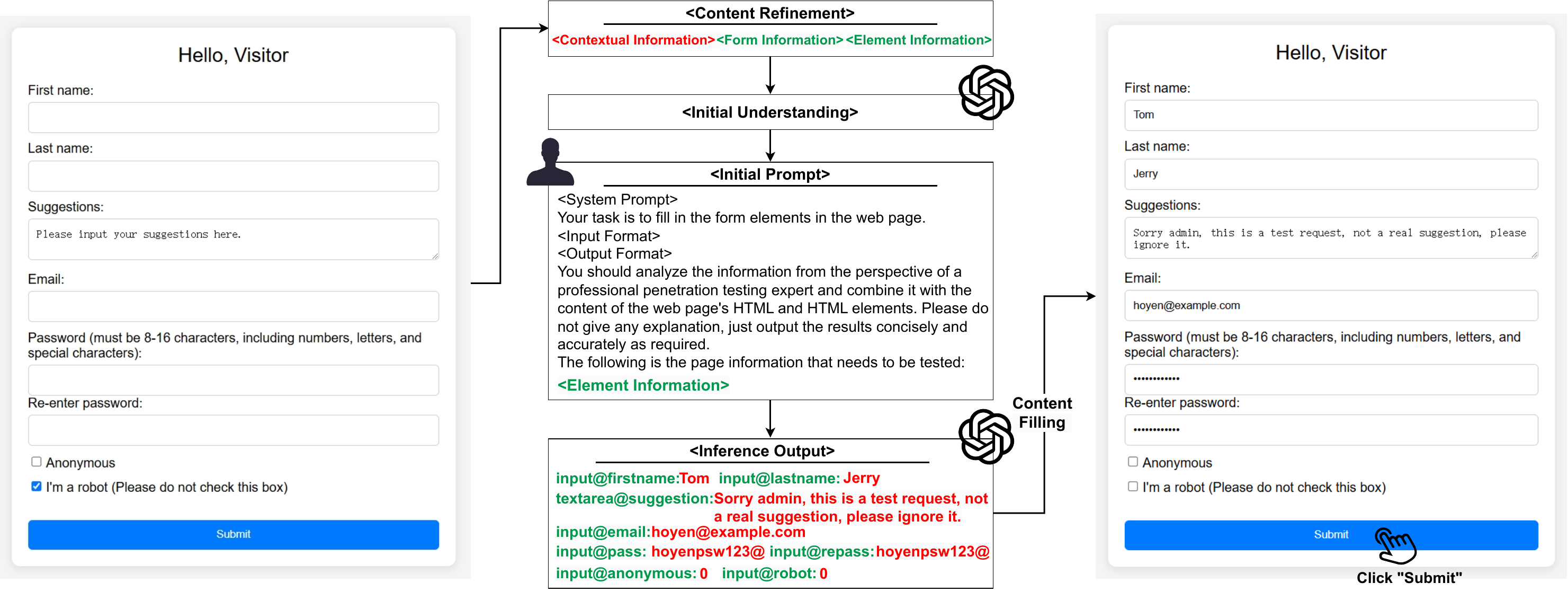}
    \caption{Example of user intention generation to the web application.}
    \label{fig:form_demo}
\end{figure*}

After identifying the user intention path, further interactions with the application’s elements are needed for in-depth exploration.
In real-world web applications, different elements offer various interaction methods, and often with dependencies between them. For instance, in Figure~\ref{fig:form_demo}, 
On the left web page, if we check the ``Anonymous" element, the ``First name" and ``Last name" will be invalid, demonstrating their dependence.
Operations on an element typically include setting values, triggering events, etc., and are closely tied to the element's semantics and context.
From the user's perspective, inter-element dependencies can be understood through contextual content, enabling the generation of operation logic based on their intent. In light of this, we propose an approach for user intention generation, which combines contextual information with user operations, enabling complex application interactions, to address complex inter-element dependencies.

Our approach uses LLM to understand user intent by considering prior operations and the overall application semantics, guiding decisions for subsequent actions. 
In particular, we would select appropriate contextual information for each element, prompts are then generated using contextual information and page content or request scenarios. During element operations, previously filled content, context, and user intention path are passed to the LLM, which integrates them to complete the tasks. To ensure consistency, prompts for all form elements are constructed uniformly and submitted to the LLM, enabling holistic processing. For nested elements, a tree-based approach is applied to perform a recursive analysis of the content within each layer to ensure precise operations. Specifically, the entire form is inserted into the prompt firstly. Then, each element that requires individual filling is extracted and analyzed, proceeding sequentially through the form. Each element's contextual information is examined individually, and the LLM follows the user intent to generate the content needed for interaction. Finally, the operations same as user intention are performed on the Web application.

During the test, \tool\, would observe and check the state changes in the web application and trace back to check if user intention and related actions are in the right direction. This process will continue until all events and elements are tested. Finally, the taint will be propagated along the application state paths to determine if there may be vulnerabilities in the application. 
The LLM emulates real user behavior by user intention generation, and minimizes erroneous element access or state modifications, allowing the scanner to understand inter-element dependencies and conduct comprehensive and accurate testing of the target web application.

\section{Evaluation}\label{section:Experiment}
\subsection{Experimental Settings}

We implemented \tool~by Python which consists of over 5,000 lines\footnote{Further details can be found on our website~\toollink}.
To evaluate the coverage extent of the scanner, we gather comprehensive access data on the server side and perform comparative analyses with other scanners.

XSS (cross-site scripting) exploits malicious HTML and JavaScript injection, requiring precise interactions with web applications and constructing an adequate exploit chain to trigger vulnerabilities. Since web application operations are highly sequential and rely on dynamic feedback, effective vulnerability exploitation methods demand real-time adjustments—a challenge for scanners but intuitive for human testers. Given the representativeness of XSS, we focused our verification on XSS vulnerabilities. To assess scanner effectiveness,  we reviewed vulnerability reports from various scanners for the same web application and manually reproduced the reported vulnerabilities for validation.

\textbf{Approaches under comparison.} We compare our scanner, \tool, with the following scanners, which are widely used in both academic research and industry: \cite{eriksson2021black}, \cite{zheng2021automatic}, \cite{w3af}, \cite{zap}, \cite{burpsuite}, and \cite{xray}. The target web application was scanned in a consistent environment using the default configurations of each scanner.

\textbf{Benchmark.} We tested 12 widely used open-source web applications, each with over 100 GitHub stars and an average of 15,000 stars, reflecting real-world adoption. The applications were deployed on separate Docker containers to ensure isolation and consistency in testing results.

\textbf{Large Language Models.} Our approach focuses on proposing a method rather than comparing LLMs performance. We tested various LLMs, including ChatGPT \cite{chatgpt} and Gemini \cite{gemini}, and found that even basic LLMs meeting the criteria significantly improve efficiency compared to traditional methods. Thus, we do not detail the impact of specific LLMs. For subsequent experiments, we primarily use the Google Gemini 1.0 API as the main LLM driver.

\subsection{Coverage}

\begin{figure*}[!htbp]
    \centering
    \includegraphics[width=\linewidth]{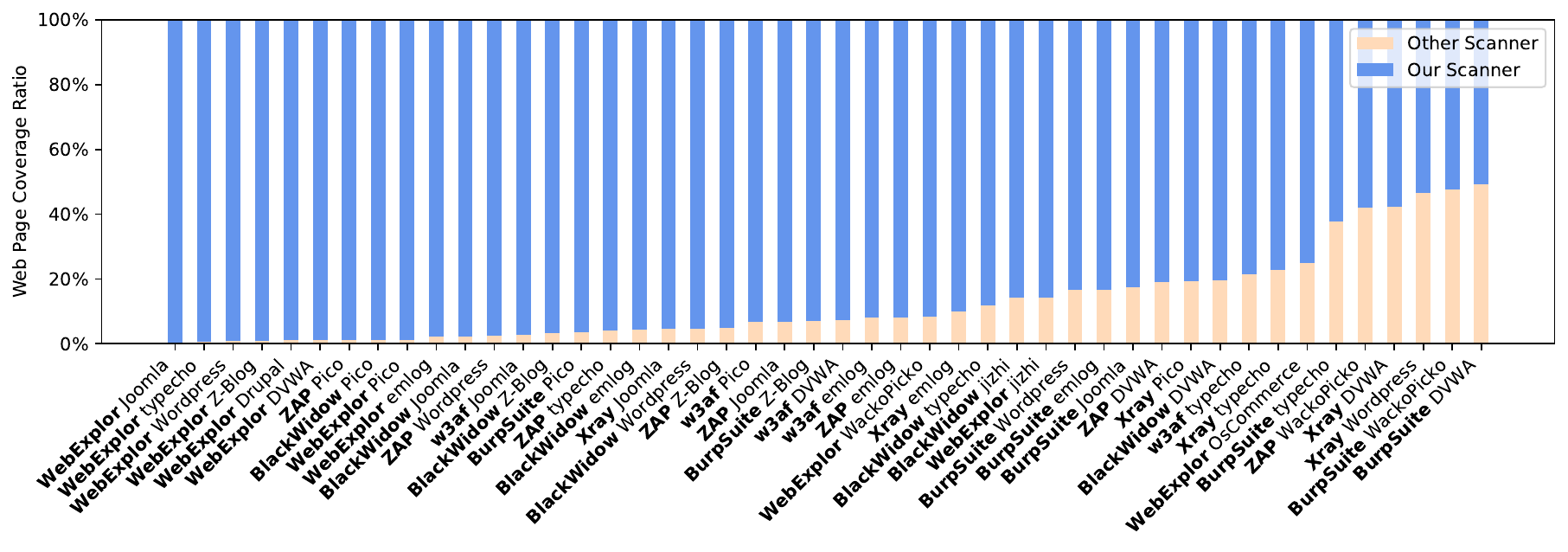}
    \caption{A comparison of page coverage between our scanner, \tool, and other scanners. Each bar is divided into two segments: one showing the page coverage achieved by \tool, and the other showing the coverage achieved by the other scanner for the same web application.}
    \label{fig:page_coverage}
\end{figure*}

As shown in Figure~\ref{fig:page_coverage}, the comparison of page coverage across different scanners for various web applications reveals that \tool, exhibits a substantial improvement in page coverage compared to other scanners, with some applications showing over a 100\% increase in coverage. The coverage results from the comparative analysis of our scanner, \tool, alongside BurpSuite and BlackWidow—currently among the most popular and state-of-the-art tools in both industry and academia are presented in Table~\ref{table:coverage_hoyen_bw}.

\begin{table}[!ht]
    \centering
    \begin{tabular}{cccc}
    \hline
        \textbf{} & \textbf{Hoyen} & \textbf{BurpSuite} & \textbf{BlackWidow } \\ \hline
        WordPress & \cellcolor{red!50}121 & \cellcolor{red!40}24 & \cellcolor{red!10}6  \\ 
        OsCommerce & \cellcolor{red!30}3 & \cellcolor{red!30}3 & \cellcolor{red!30}3  \\ 
        Prestashop & \cellcolor{red!25}1 & \cellcolor{red!50}{8} & \cellcolor{red!25}1  \\ 
        Joomla & \cellcolor{red!50}222 & \cellcolor{red!40}47 & \cellcolor{red!10}5  \\ 
        WackoPicko & \cellcolor{red!30}11 & \cellcolor{red!20}8 & \cellcolor{red!50}13  \\ 
        Drupal & \cellcolor{red!25}93 & \cellcolor{red!25}98 & \cellcolor{red!50}198  \\ 
        Pico & \cellcolor{red!50}84 & \cellcolor{red!25}3 & \cellcolor{red!25}1  \\ 
        typecho & \cellcolor{red!50}143 & \cellcolor{red!40}87 & \cellcolor{red!10}19  \\ 
        emlog & \cellcolor{red!50}45 & \cellcolor{red!30}9 & \cellcolor{red!20}2 \\
        Z-Blog & \cellcolor{red!50}119 & \cellcolor{red!30}9 & \cellcolor{red!20}4 \\
        jizhi & \cellcolor{red!40}6 & \cellcolor{red!50}23 & \cellcolor{red!10}1 \\
        DVWA & \cellcolor{red!50}90 & \cellcolor{red!30}87 & \cellcolor{red!20}22  \\ 
        \hline
    \end{tabular}
    \caption{The page number covered by \tool, BurpSuite, and BlackWidow across various web applications on the server.}
    \label{table:coverage_hoyen_bw}
\end{table}
\begin{table}[!ht]
    \centering
    \begin{tabular}{cccc}
    \hline
        \textbf{} & \textbf{Hoyen} & \textbf{BurpSuite} & \textbf{BlackWidow} \\ \hline
        Wordpress & \cellcolor{red!50}0.99 & 0.78 & \cellcolor{red!30}0.95 \\ 
        OsCommerce & \cellcolor{red!50}1.00 & 0.70 & \cellcolor{red!50}1.00 \\ 
        Prestashop & \cellcolor{red!50}1.00 & \cellcolor{red!30}0.88 & \cellcolor{red!50}1.00 \\ 
        Joomla & \cellcolor{red!50}1.00 & 0.73 & \cellcolor{red!50}1.00 \\ 
        WackoPicko & \cellcolor{red!50}0.99 & 0.72 & 0.61 \\ 
        Drupal & \cellcolor{red!50}1.00 & 0.64 & \cellcolor{red!50}1.00 \\ 
        Pico & \cellcolor{red!50}1.00 & 0.30 & \cellcolor{red!30}0.98 \\ 
        typecho & \cellcolor{red!50}1.00 & \cellcolor{red!30}0.84 & \cellcolor{red!50}1.00 \\ 
        emlog & \cellcolor{red!50}1.00 & 0.67 & \cellcolor{red!50}1.00 \\ 
        Z-Blog & \cellcolor{red!40}0.94 & \cellcolor{red!30}0.93 & \cellcolor{red!50}0.99 \\ 
        DVWA  & \cellcolor{red!40}0.88  & \cellcolor{red!30}0.86  & \cellcolor{red!50}0.99  \\ \hline
    \end{tabular}
    \caption{The rate of successful requests for \tool, BurpSuite, and BlackWidow to all their requests.}
    \label{table:req_success}
\end{table}
Using BlackWidow as a baseline, Hoyen demonstrated comparable or superior performance in 10 out of 12 web applications, achieving a 100\% increase in overall average page coverage. For globally popular platforms like WordPress and Joomla \cite{w3techCMS}, Hoyen identified 20 times more new pages than BlackWidow. On smaller applications like DVWA, Hoyen achieved a four-fold improvement in page coverage. Similarly, compared to BurpSuite, Hoyen outperformed in 8 out of 12 web applications.
BlackWidow excelled on WackoPicko and Drupal due to its conservative page similarity merging strategy, leading to repeated scans of similar pages. For example, in our Drupal environment with a language-switching feature, BlackWidow rescanned nearly identical pages in different languages, achieving more page coverage. In contrast, our approach avoids redundant scans by filtering similar pages based on comparisons. On WackoPicko, BlackWidow tested uploaded file content extensively, whereas our approach prioritized XSS vulnerability detection, placing less emphasis on file upload functions.  BurpSuite outperformed on PrestaShop, leveraging a dictionary-based method to discover sensitive directories and access additional pages. Our approach, lacking dictionary or brute-force techniques, may miss some web application entry points due to incomplete target system information.

\begin{table*}[!t]
    \centering
    \begin{tabular}{cccccccc}
    \hline
    {}&\multicolumn{4}{c}{\textbf{General functions}} & \multicolumn{2}{c}{\textbf{Core functions}} & \multirow{2}*{\textbf{Others}} \\
    \cmidrule(lr){2-5} \cmidrule(lr){6-7} 
        {} & \textbf{Homepage} & \textbf{Login} & \textbf{Setup} & \textbf{About} & \textbf{Vulnerable} & \textbf{Normal} & ~ \\ \hline
        \textbf{Hoyen} & 1.40\% & 2.00\% & 0.00\% & 1.60\% & \cellcolor{red!50}58.90\% & \cellcolor{red!30}36.00\% & 0.10\%  \\ 
        \textbf{w3af} & \cellcolor{red!10}10.30\% & \cellcolor{red!20}24.40\% & 2.60\% & 1.30\% & 0.00\% & 0.00\% & \cellcolor{red!60}61.50\%  \\ 
        \textbf{Xray} & 9.40\% & \cellcolor{red!10}11.80\% & 2.30\% & 2.20\% & 1.50\% &\cellcolor{red!30} 35.30\% & \cellcolor{red!30}37.50\%  \\ 
        \textbf{ZAP} & 2.40\% & \cellcolor{red!40}40.50\% & 2.40\% & 2.40\% & 9.60\% & \cellcolor{red!40}40.70\% & 2.40\%  \\ 
        \textbf{BurpSuite} & 6.00\% & 9.30\% & 3.10\% & 0.50\% & \cellcolor{red!10}17.70\% & \cellcolor{red!60}60.60\% & 2.70\%  \\ 
        \textbf{BlackWidow} & 1.50\% & \cellcolor{red!60}65.20\% & 3.90\% & 2.80\% & \cellcolor{red!10}12.00\% & \cellcolor{red!10}14.60\% & 0.00\%  \\ 
        \hline
    \end{tabular}
    \caption{The table of requests from various scanners regarding the different functionality of DVWA. Each cell represents the percentage of all requests for this function.}
    \label{table:heat_dvwa_fun}
\end{table*}
\subsection{Effectivity}

Scanners employ varying strategies for path detection, aiming to access as many web application pages as possible for comprehensive testing. Standard methods typically start at the home page and analyze links sequentially or randomly. Some scanners may use dictionary-based brute force or enumeration, which can disrupt normal service operations. While academic research avoids brute force, scanners often generate numerous invalid requests that deviate from human navigation patterns \cite{jueckstock2021towards,li2023scan}. To address this, we incorporate request accuracy into our evaluation, ensuring the scanner focuses on critical functions while maintaining high accuracy. Higher error rates in requests often distinguish scanners from human users \cite{li2023scan}.

As shown in Table~\ref{table:req_success}, our approach achieves nearly 100\% successful request rates. This demonstrates that our requests efficiently access web application functions. 
Our approach to vulnerability detection in web applications follows a strategy similar to BlackWidow's, starting with small-scale testing during web application modeling and crawling, followed by comprehensive vulnerability assessment at identified edges. Both approaches generate fewer erroneous requests compared to other scanners. In our case, erroneous requests primarily occur during vulnerability testing. This is due to our broader efforts to identify additional vulnerabilities, which require more testing requests. The higher number of erroneous requests in DVWA and Z-Blog is linked to detecting more pages than BlackWidow, resulting in more erroneous requests of vulnerability tests. Despite this, our scanning strategy maintains high accuracy.

As shown in Table~\ref{table:heat_dvwa_fun}, different scanners prioritize different aspects of web applications. For example, w3af only focused on a small portion of the core functionality. BlackWidow tested core functionality and identified vulnerabilities but concentrated significantly on the login section, potentially diverting attention from the core functions. BurpSuite effectively allocated 78\% of its efforts to core functions, but its scans were more dispersed due to limited page comprehension. Our approach, emphasizing XSS vulnerability detection, dedicated over 30\% of requests to XSS tests, and an unexpected function detected an XSS vulnerability, accounting for nearly 25\% of the total request attention. Overall, our approach proved highly effective, focusing nearly 95\% of its attention on the web application's core functionality, which was key to vulnerability identification. Based on experimental verification, we identified more vulnerabilities than other approaches. In addition to the vulnerabilities detected by existing tools, we discovered 4 unique vulnerabilities.

\section{Conclusion}

In response to the limitations of contemporary black-box web application scanners, this paper introduces \tool, a novel contextual semantics and content awareness black-box web application scanning approach driven by the LLMs. The approach involves identifying and selecting elements on web pages through a function-based page content refinement approach, and semantic and content-awareness-based user intention prediction approach. These approaches allow a deeper understanding of the web application's structure and elements, enabling a more comprehensive application analysis. In our experiments, \tool\, demonstrated a $2\times$ increase in average page coverage. The test requests were precisely directed toward the core functions of web applications, and identifying more vulnerabilities.
Across all performance metrics, \tool\, outperformed existing black-box scanners in web applications.

\appendix

\section*{Ethical Statement}

To address potential legal and ethical concerns associated with the experiment on web scanners, we referenced the scanning red lines outlined in \cite{hantke2024red}, constructed the target system in our laboratory, and conducted comparative experiments involving our scanner and other related works. Additionally, we followed ethical guidance from \cite{demir2022reproducibility} to ensure and regulate the proper conduct of our scanner. Our scanner also appends a unique tag to all User-Agent requests for detection monitoring on the server side. Throughout the evaluation process, we ensured that our experiments had minimal impact on the system. To maintain consistency in experimental results \cite{jueckstock2021towards}, all tests were performed on servers located in our laboratory, with all test environments locally deployed.

\bibliographystyle{named}
\bibliography{reference}

\end{document}